\documentclass{iopart}

\usepackage{graphics}
\usepackage{graphicx}
\usepackage{graphics}
\usepackage{iopams} 
\usepackage{setstack} 
\usepackage{upgreek}
\usepackage[colorlinks,citecolor=blue,urlcolor=blue,linkcolor=blue]{hyperref}

\newcommand{\gerda}     {{\sc Gerda}}

\newcommand{\igex}     {{\sc Igex}}
\newcommand{\hdm}     {{\sc HdM}}

\def\twonu {$2\nu \beta\beta$}

\def\gess      {{$^{76}$Ge}}

\def\thalf {T^{2\nu}_{1/2}}
\newcommand{\geenr}       {{$^{\rm enr}$Ge}}          
\newcommand{\genat}       {{$^{\rm nat}$Ge}}

\begin{document}
\title[Measurement of the half-life of the two-neutrino 
$\beta\beta$ decay of $^{76}$Ge with \gerda]{Measurement of the half-life of the two-neutrino 
double beta decay of $^{76}$Ge with the \gerda\ experiment}


\author{The \gerda\ Collaboration \\
   M~Agostini$^{14}$,  
   M~Allardt$^{3}$,  
   E~Andreotti$^{5,18}$,  
   A~M~Bakalyarov$^{12}$,  
   M~Balata$^{1}$,  
   I~Barabanov$^{10}$,  
   M~Barnab\'e~Heider$^{14}$\footnote{\emph{Present Address:} CEGEP St-Hyacinthe, 
    Qu{\'e}bec, Canada},  
   N~Barros$^{3}$,  
   L~Baudis$^{19}$,  
   C~Bauer$^{6}$
% ...
}

\address{$^1$ INFN Laboratori Nazionali del Gran Sasso, LNGS, Assergi, Italy}   
\address{$^2$ Institute of Physics, Jagiellonian University, Cracow, Poland}   
\address{$^3$ Institut f{\"u}r Kern- und Teilchenphysik, Technische Universit{\"a}t Dresden,
      Dresden, Germany}
\address{$^4$ Joint Institute for Nuclear Research, Dubna, Russia}
\address{$^5$ Institute for Reference Materials and Measurements, Geel,
     Belgium}
\address{$^6$ Max-Planck-Institut f{\"u}r Kernphysik, Heidelberg, Germany}   
\address{$^7$ Dipartimento di Fisica, Universit{\`a} Milano Bicocca, 
     Milano, Italy}
\address{$^8$ INFN Milano Bicocca, Milano, Italy} 
\address{$^9$ Dipartimento di Fisica, Universit{\`a} degli Studi di Milano e INFN Milano,
    Milano, Italy}
\address{$^{10}$ Institute for Nuclear Research of the Russian Academy of Sciences,
    Moscow, Russia}
\address{$^{11}$ Institute for Theoretical and Experimental Physics,
    Moscow, Russia}
\address{$^{12}$ National Research Centre ``Kurchatov Institute'', Moscow, Russia}
\address{$^{13}$ Max-Planck-Institut f{\"ur} Physik, M{\"u}nchen, Germany}
\address{$^{14}$ Physik Department and Excellence Cluster Universe,
    Technische  Universit{\"a}t M{\"u}nchen, Germany}
\address{$^{15}$ Dipartimento di Fisica e Astronomia dell'Universit{\`a} di Padova, 
    Padova, Italy}
\address{$^{16}$ INFN  Padova, Padova, Italy}
\address{$^{17}$ Shanghai Jiaotong University, Shanghai, China}
\address{$^{18}$ Physikalisches Institut, Eberhard Karls Universit{\"a}t T{\"u}bingen,
    T{\"u}bingen, Germany}
\address{$^{19}$ Physik Institut der Universit{\"a}t Z{\"u}rich, Z{\"u}rich, 
    Switzerland}
\ead{pandola@lngs.infn.it}

\begin{abstract}
The primary goal of the GERmanium Detector Array (\gerda) experiment at 
the Laboratori Nazionali del Gran Sasso of INFN is the search for the 
neutrinoless double beta decay of $^{76}$Ge. 
High-purity germanium detectors made from material enriched 
in $^{76}$Ge are operated directly immersed in liquid argon, 
allowing for a substantial reduction of 
the background with respect to predecessor experiments. 
The first 5.04~kg$\cdot$yr of data collected in Phase~I of the experiment 
have been analyzed to measure the half-life of the neutrino-accompanied 
double beta decay of $^{76}$Ge. 
The observed spectrum in the energy range between 600 and 1800~keV is 
dominated by the double beta decay of $^{76}$Ge. 
The half-life extracted from \gerda\ data is 
$\thalf$ = $(1.84 ^{+0.14}_{-0.10}) \cdot 10^{21}$ yr.

\end{abstract}
\submitto{\jpg}
\maketitle
%
\section{Introduction and scope} \label{sec:intro}

Neutrinoless double beta ($0\nu\beta\beta$) decay of atomic nuclei 
($A$,$Z$) $\rightarrow$ ($A$,$Z + 2$) +2$e^{-}$ is a forbidden process  
in the Standard Model (SM) of particle physics 
because it violates lepton number by two units. 
An observation of such a decay would demonstrate lepton number 
violation in nature and would prove that neutrinos have a Majorana 
component. For recent reviews, see \cite{reviews}.  
The two-neutrino double beta ($2\nu\beta\beta$) decay of atomic nuclei, 
\begin{displaymath}
(A,Z) \rightarrow (A,Z + 2) + 2 e^{-} + 2\overset{-}{\nu}_e, 
\end{displaymath}
with the simultaneous emission of two electrons and two anti-neutrinos, 
conserves lepton number and 
is allowed within the SM, independent of the nature of 
the neutrino. Being a higher-order process, it is characterized by an 
extremely low decay rate: so far it is the rarest decay observed in 
laboratory experiments. It is observable for a few even-even 
nuclei and was detected to-date for eleven  
nuclides; the corresponding half-lives are in the range of 
$7 \cdot 10^{18} - 2 \cdot 10^{24}$~yr \cite{bar10,tre95,tre02}. \\

The measurement of the half-life of the \twonu\ decay ($\thalf$) is of 
substantial interest. For example, model
predictions of the $0\nu\beta\beta$ half-life require the evaluation of nuclear
matrix elements. These calculations are complicated and have large 
uncertainties. They are different from those required for 
the \twonu\ decay, but it has been 
suggested~\cite{rod06,sim} 
that, within the same model framework, some constraints on the 
$0\nu\beta\beta$ matrix elements  NME$^{0\nu}$ can be derived from the 
knowledge of the \twonu\ nuclear matrix elements NME$^{2\nu}$. Also, 
the nuclear matrix element NME$^{2\nu}$ which is extracted from the measurement 
of the half-life of the \twonu\ decay can be directly compared with the 
predictions based on charge exchange experiments~\cite{kvi,rcnp}. A 
good agreement would indicate that the reaction mechanisms and the 
nuclear structure aspects that are involved in the \twonu\ 
decay are well understood. \\

The GERmanium Detector Array (\gerda) experiment at the Laboratori 
Nazionali del Gran Sasso of INFN searches for the $0\nu\beta\beta$ decay 
of $^{76}$Ge. 
High-purity germanium detectors isotopically enriched in \gess\ are operated 
bare and immersed in liquid argon in order to greatly reduce the 
environmental background. As a first result of this ongoing research, the present paper 
reports a precise measurement of the half-life of the \twonu\ decay 
of $^{76}$Ge. The data used in this work encompasses an 
exposure of 5.04~kg$\cdot$yr, taken between November 2011 and 
March 2012.\\

\section{The \gerda\ experimental setup} \label{sec:setup}

\begin{figure}[tpb]
\begin{center}
\includegraphics[width=0.65\columnwidth]{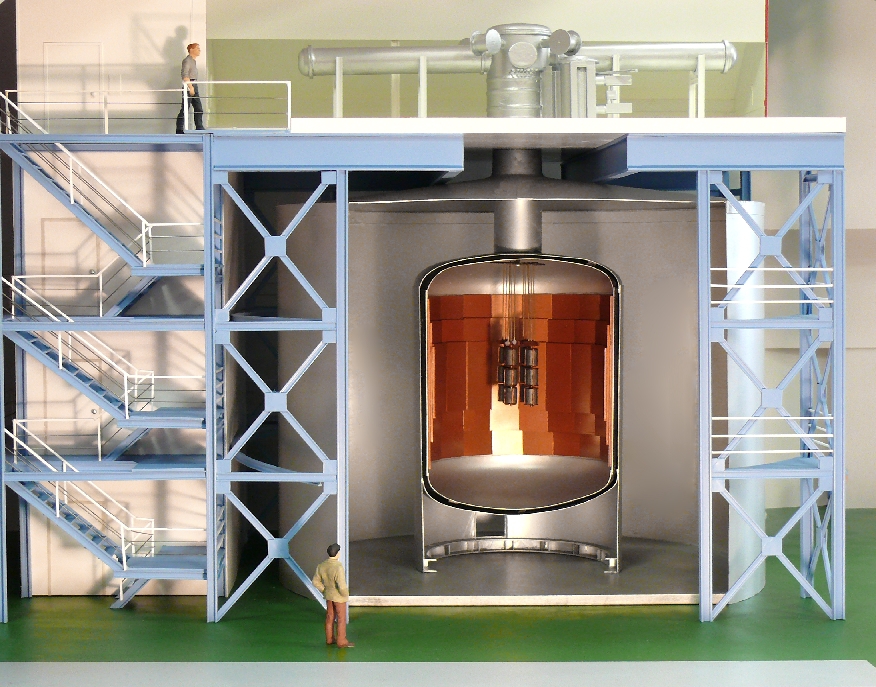}
\caption{Artist's view of the \gerda\ experiment. The detector array is not to scale.}\label{fig:setup}
\end{center}
\end{figure}
A brief outline of the components of the \gerda\ detector that are most relevant for this 
work is given below; a detailed description can be found in~\cite{Gerda,NIM_paper}. \\

The \gerda\ experimental setup is shown in figure ~\ref{fig:setup}. At the core of the 
setup there is an array of high-purity germanium detectors (HPGe). They are operated bare 
in liquid argon (LAr) which acts both as a coolant and as a shield against the residual 
environmental background. 
The array configuration consists of eleven germanium detectors: eight 
are made from isotopically modified germanium (\geenr), enriched to about 86\% in \gess, 
and three are made from natural germanium (\genat), with a total mass of 17.67~kg 
and 7.59~kg, respectively. The enriched detectors come from the former Heidelberg-Moscow 
(\hdm)~\cite{gun97} and \igex~\cite{igex99} experiments.  
They underwent specific refurbishing processes before operation in \gerda~\cite{MB_PhD,IEEE}. 
The germanium detectors 
are mounted in strings with typically three diodes each. Signals are amplified by  
low noise, low radioactivity charge sensitive preamplifiers~\cite{ribol} with 30~MHz bandwidth 
operated inside the LAr. They are digitized by a 14-bit 100~MHz 
continuously running ADC (FADC) equipped with anti-aliasing bandwidth filters.
In the offline analysis the waveforms are digitally 
processed to reconstruct the event energy.\\

The detector array is surrounded by 64 m$^3$ of 5.0-grade LAr, contained in 
a vacuum insulated cryostat made of stainless steel, lined on the inner side by a 3 to 6~cm 
thick layer of copper. 
The cryostat is in turn placed at the centre of a 580~m$^3$ volume of ultra-pure water 
equipped with 66 photomultiplier tubes to veto the residual cosmic ray muons   
by the detection of Cherenkov light. The large water volume also serves as a shield 
to moderate and capture neutrons produced by natural 
radioactivity and in muon-induced hadronic showers.\\  

The energy scale is set by using calibration curves, parametrized as second-order 
polynomials, derived for each detector by calibration runs taken with $^{228}$Th sources. 
The stability of the energy scale is monitored by performing such   
calibration runs every one or two weeks. Moreover, the stability of the system is 
continuously monitored by using ad hoc charge pulses generated by a spectroscopy 
pulser that are regularly injected in the input of the charge sensitive 
preamplifier.\\
 
All \gerda\ detectors but two exhibit a reverse current of the order of tens of pA. The 
two problematic detectors showed an increase 
of leakage current soon after the beginning of their operation; therefore their 
bias high voltage had to be reduced and finally completely removed. These two detectors 
do not contribute to the present data set, although they are accounted for in the 
detector anti-coincidence cut described below. The total mass of the operational enriched 
detectors is 14.63~kg. The \genat\ detectors are not considered in the present 
analysis because of their low content of \gess. 
The average energy resolution for the six enriched detectors considered 
in this study is about 4.5~keV (full width at half maximum) at the 
Q$_{\beta\beta}$-value of $^{76}$Ge (2039~keV)~\cite{NIM_paper}.\\
 
A key parameter which is required for the computation of $\thalf$ is the total $^{76}$Ge 
active mass. It is calculated as the product of the total detector mass with the isotopic 
abundance of \gess\ ($f_{76}$) and the fraction of the active mass ($f_{act}$).\\ 
The average \gess\ isotopic abundance of the six enriched detectors 
considered in this work is 86.3\%~\cite{NIM_paper,igex99,igex03,klap01}.\\ 
In \gerda\ p-type semi-coaxial detectors are used, for which a part of the volume 
close to the outer surface is inactive.
After mechanical modifications and processing of the germanium diodes at the manufacturer, 
their active mass was re-determined experimentally~\cite{NIM_paper,MB_PhD}.  The measured average 
fraction $f_{act}$ is 86.7\%, with individual detector $f_{act}$ uncertainties 
of about 6.5$\%$. 

\section{The data set} \label{sec:dataset}
The data set considered for the analysis was taken 
between November 9, 2011, and March 21, 2012, for a total of 125.9~live 
days, amounting to an exposure of 5.04~kg$\cdot$yr. Digitized charge pulses from the 
detectors are analyzed with the software 
tool \textsc{Gelatio}~\cite{gelatio} according to the \gerda\ standard 
procedure~\cite{acat}. 
The pulse amplitude is reconstructed offline by applying an approximated Gaussian 
filter with an integration time of 5~$\mu$s (which is sufficient to avoid losses due to 
ballistic effects). 
Events generated by discharges or due to electromagnetic noise are 
rejected by using a set of 
quality cuts.
Due to the low counting rate, 
the data set has a negligible contamination of pile-up events. The 
combined efficiency of the trigger and of the  
offline data processing is practically 100\% above 100~keV. Similarly, no loss of 
physical events is expected from the application of the quality cuts, as 
deduced by dedicated Monte Carlo studies and by the analysis of pulser data. 
Events that are in coincidence with a valid veto signal from the muon detector and 
events having a signal in more than one HPGe detector are excluded 
from the analysis. The time window for the coincidence between the muon 
detector and the HPGe detectors was set to 8~$\mu$s while that between different 
HPGe detectors was set to a few $\mu$s. Due to the very low muon flux 
in the Gran Sasso laboratory, the dead time induced by the 
muon veto is negligible. \\

\begin{figure}[tpb]
\begin{center}
\includegraphics[width=0.85\columnwidth]{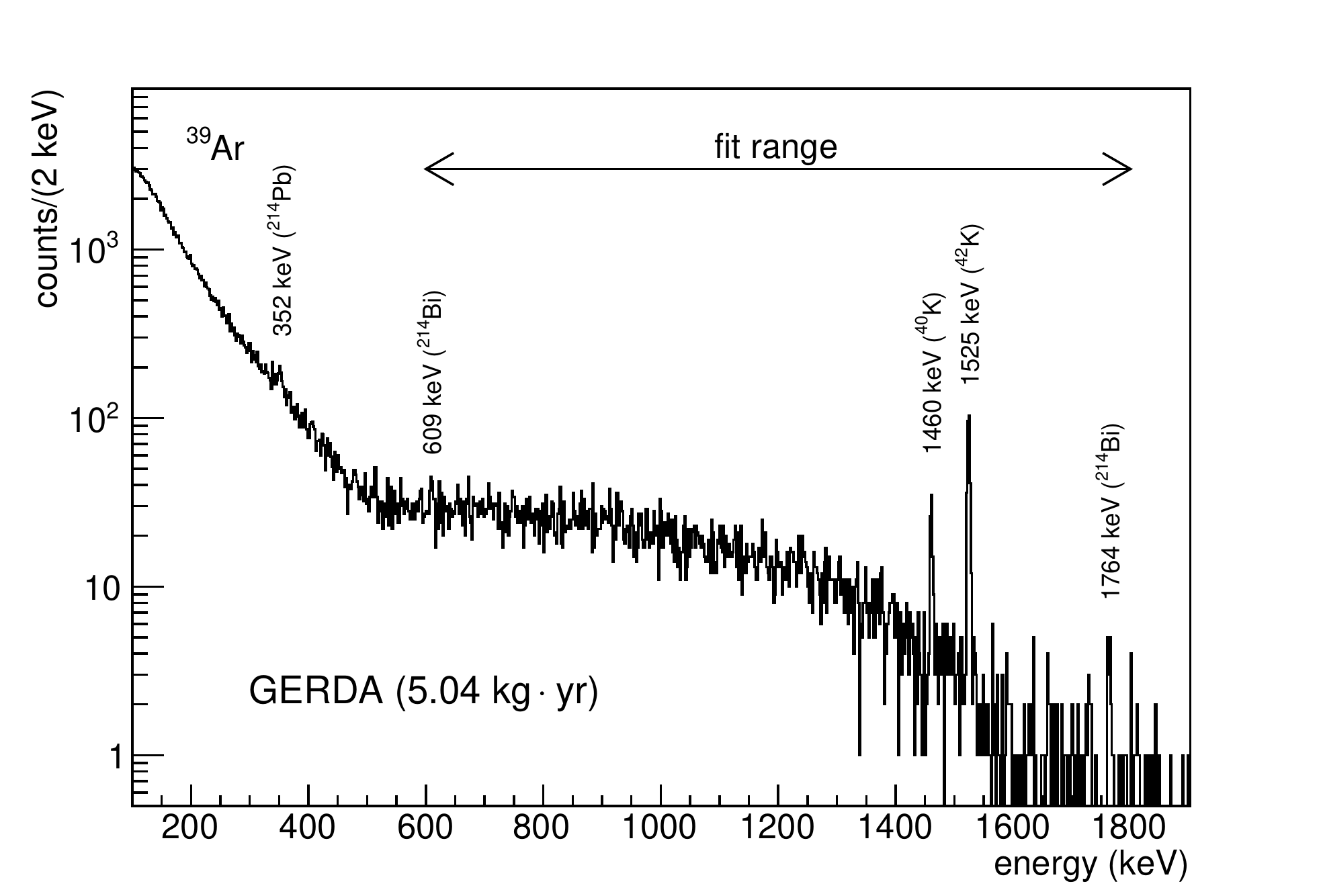}
\caption{Energy spectrum of the six \geenr\ detectors. The $^{39}$Ar 
$\beta$ decay dominates the counting rate at low energies up to its 
$Q_{\beta}$-value of 565~keV. The energy range which is considered 
for the \twonu\ analysis and a few prominent $\gamma$ lines are 
also shown.} \label{fig:spectrum}
\end{center}
\end{figure}

Given the half-life of the \twonu\ decay reported in the literature 
(about 1.5 $\cdot 10^{21}$~yr), 
the anticipated count rate of the \geenr\ detectors is about 100~counts/day in the entire 
energy range up to $Q_{\beta\beta}$=2039~keV. Since the detectors are submerged 
in LAr, the radioactive decay of $^{39}$Ar, which is a long-lived $\beta$ 
emitter produced by cosmogenic activation of natural argon in the atmosphere, gives a 
large contribution up to its $Q_{\beta}$-value of 565~keV.  In fact, the low-energy spectrum 
is dominated by these $\beta$ particles and their Bremsstrahlung photons, 
which account for about 1000~counts/day 
above 100~keV. The \twonu\ decay is expected to be the major contributor in the energy spectrum 
above the end-point of the $^{39}$Ar spectrum.
For this reason, the analysis of the \twonu\ decay is performed in the range between 
600~keV, which is comfortably above the end-point of the $^{39}$Ar spectrum,  
and 1800~keV. The sum spectrum of the 
six \geenr\ detectors considered in this work is displayed in figure~\ref{fig:spectrum}.  
The analysis range contains 8796~events in total. 
The probability for a \twonu\ decay taking place in the active volume of the \geenr\   
detectors to produce a total energy release between 600 and 1800~keV is about 63.5\%; 
the energy range above 1800~keV is practically insensitive 
to the \twonu\ signal, as the probability for a decay to produce an energy release in this 
region is $< 0.02\%$. These estimates are based on the Monte Carlo simulation of 
\twonu\ decays in the \gerda\ detectors as described in section \ref{sec:analysis}.
Hence, the energy region chosen for the analysis is well suited for 
the study of the \twonu\ decay signal.
\section{Data analysis} \label{sec:analysis} 
\subsection{Statistical treatment and fit model}

The experimental spectra of the six diodes are analyzed following 
the binned maximum likelihood approach described in \cite{allen}. The analysis 
region is divided into 40 bins, each 30~keV wide. A global model 
is fitted to the observed energy spectra. The model contains
the \twonu\ decay of $^{76}$Ge and 
three independent background contributions, namely $^{42}$K, $^{214}$Bi and 
$^{40}$K. The presence of these background sources is established by 
the observation of their characteristic $\gamma$ lines: 1525~keV from 
$^{42}$K; 1460~keV from $^{40}$K; 609~keV and 1764~keV from $^{214}$Bi; 
352~keV from $^{214}$Pb (progenitor of $^{214}$Bi). \\

$^{42}$K is a short-lived $\beta$ emitter ($Q_{\beta}$= 3525 keV,
$T_{1/2}$= 12.6 h) which is present in \gerda\ as a progeny of the long-lived
$^{42}$Ar ($Q_{\beta}$= 599 keV, $T_{1/2}$= 32.9 yr). $^{42}$Ar is a trace
radioactive contaminant expected in natural argon and is produced by cosmogenic
activation. $^{214}$Bi from the $^{238}$U decay series and $^{40}$K are
$\gamma$ emitters from the environmental radioactivity.  \\
A few more candidate $\gamma$ lines have been identified in the \gerda\ 
spectrum~\cite{NIM_paper}, which indicate the presence of small additional 
background contributions: in particular $^{208}$Tl and $^{228}$Ac 
from the $^{232}$Th decay series, and $^{60}$Co. However, most of 
the candidate $\gamma$ lines have either a poor statistical significance and/or 
are seen in some detectors only. Given the lack of discriminating power in the 
data, the background contributions other than $^{42}$K, $^{214}$Bi and $^{40}$K 
are not included in the fit. However, their possible impact on the extracted 
half-life $\thalf$ is included in the systematic uncertainty, as discussed 
in section \ref{sec:syst}; their cumulative contribution to the background 
is estimated to be of a few percent.\\

The half-life of the \twonu\ decay is common in the fit to the six spectra 
of the \geenr\ detectors.
The intensities of the background components are independent 
for each detector. The active mass and the \gess\ abundance of each detector 
are also left free in the fit; they are treated as nuisance parameters and 
integrated over at the end of the analysis. \\

The shapes of the energy spectra for the model components (signal and
three backgrounds) are derived by a Monte Carlo simulation for each detector individually.
The simulation is performed using the \textsc{MaGe} framework~\cite{mage} based on
\textsc{Geant4}~\cite{geant4}. Assumptions have to be made in the
Monte Carlo simulation about the location and the primary spectrum of each
component of the model. The spectrum of the two electrons emitted in the
\twonu\ decay of $^{76}$Ge is sampled according to the distribution   
of \cite{tre95} that is implemented in the code 
\textsc{decay0}~\cite{decay0}. Electrons are propagated in the \gerda\ simulated
setup by \textsc{MaGe} and the total energy released in the active mass of
the enriched detectors is registered.\\ 
The $^{42}$K activity is uniformly
distributed in the liquid argon volume. The decay products of the $\beta$ decay 
of $^{42}$K are taken into account as the initial state in the simulation. 
However the energy deposit in the detectors is mainly given by the 
1525 keV $\gamma$-ray (full energy peak and Compton continuum): 
the contribution due to the $\beta$ particles is small (less than a few percent) 
with respect to the 1525~keV $\gamma$-ray. The actual position of the $^{40}$K
and $^{214}$Bi emitters contributing to the \gerda\ spectrum is not 
known in detail: the assumption is made in the Monte Carlo of ``close sources''. 
The ratio of the intensities of the $^{214}$Bi $\gamma$ lines 
observed in the experimental spectrum is consistent with such an assumption.  
The impact on 
$\thalf$ due to the lack of knowledge about the source position -- which
affects the peak-to-Compton ratio -- is accounted as a systematic  
uncertainty. The effect of muon-induced events that are not accompanied by a
veto signal, e.g. due to inefficiency, is estimated to be $< 0.02\%$ in the
energy range of this analysis.\\

The spectral fit is performed using the Bayesian Analysis Toolkit 
\textsc{Bat}~\cite{bat}. A flat distribution between 0 and $10^{22}$~yr is  
taken as the prior probability density function (pdf) for $\thalf$. 
The prior pdf for the active mass fraction of each
detector is modelled as a Gaussian distribution, having mean value
and standard deviation according to the measurements performed in
\cite{MB_PhD}. The analysis accounts for the fact that the uncertainties 
on the active masses $f_{act}$ are partially correlated, because of the experimental procedure 
employed for the measurement. Uncertainties on $f_{act}$ are split into 
an uncorrelated term $-$ which is specific of each detector $-$ and a common 
correlated term. The prior pdf for the \gess\ isotopic abundance 
$f_{76}$ for each detector  
is also a Gaussian, having mean value according to the earlier measurements 
reported in \cite{NIM_paper}. The uncertainty of $f_{76}$ is 
between 1\% and 3\% for the single detectors. 
It was estimated from the dispersion of independent measurements performed with
isotopically enriched material; this dispersion is much larger than the quoted
uncertainty of each individual measurement. \\ 
The fit has 32 free parameters: one common value for the \twonu\ half-life and
31 nuisance parameters. The nuisance parameters are the common term which describes  
the correlated uncertainty of the active masses plus five independent 
variables for each of the six detectors (active mass, \gess\   
abundance and three background components). The detector parameters used for 
the prior pdf's are summarized in table~\ref{tab:detectors}. \\ 

\begin{figure}[tpb]
\begin{center}
\includegraphics[width=0.85\columnwidth]{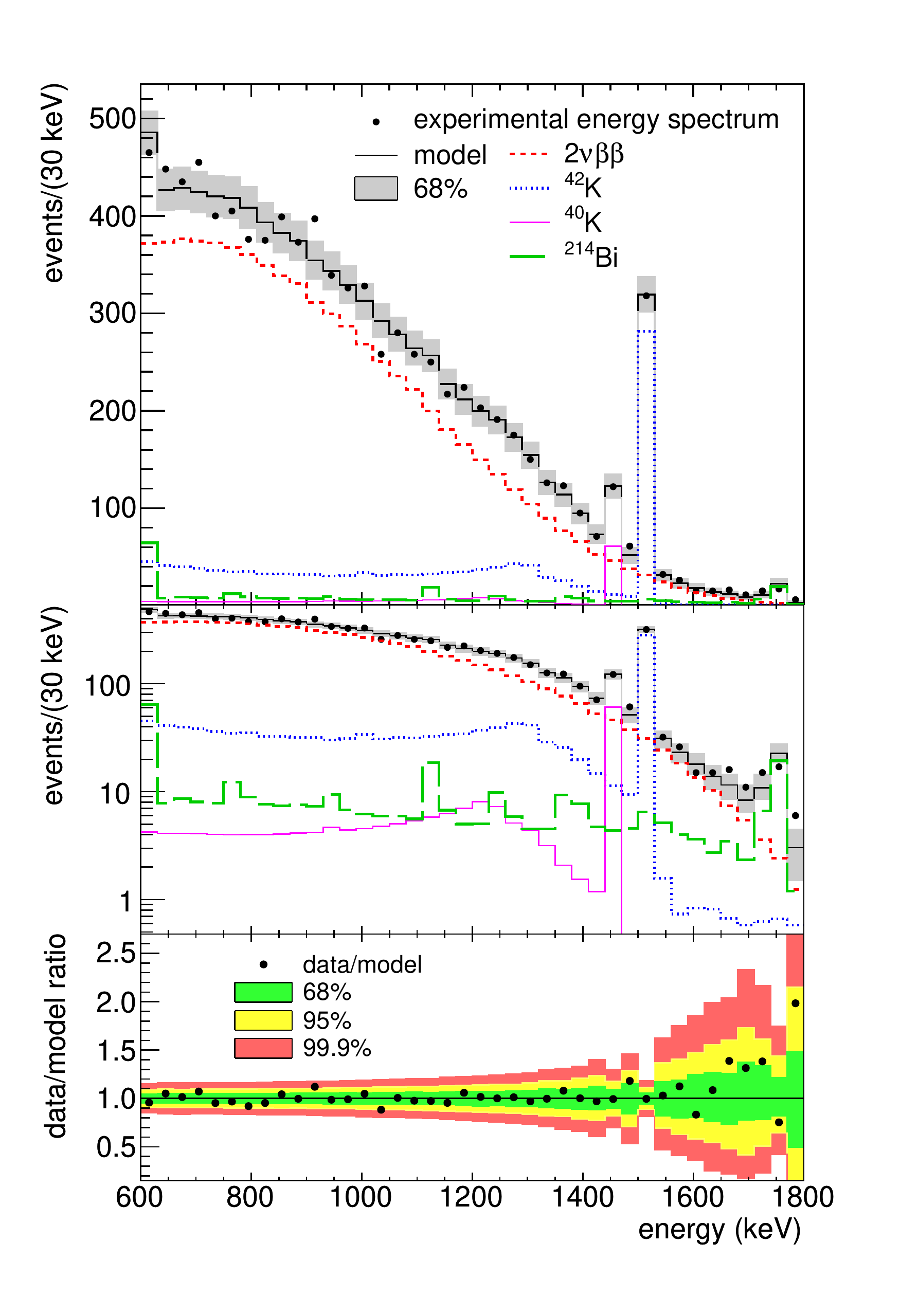}
\caption{
  Upper and middle panels: Experimental data (markers) and the 
  best fit model (black histogram) for the sum of the six detectors together 
  (linear and logarithmic scale). 
  Individual contributions from \twonu\ 
  decay (red), $^{42}$K (blue), $^{40}$K (purple) and $^{214}$Bi (green) 
  are shown separately. The shaded band covers the 68\% probability range for the 
  data calculated from the expected event counts of the best fit model.
  Lower panel: ratio between experimental data and the prediction of the best fit 
  model. The green, yellow and red regions are the smallest intervals containing 
  68\%, 95\% and 99.9\% probability for the ratio assuming the best fit 
  parameters, respectively~\cite{colorplots}.}\label{fig:model}
\end{center}
\end{figure}

Figure~\ref{fig:model} shows experimental data together with the best fit model 
for the sum of the six detectors. The analysis energy window contains 8796 events. 
The best fit model has an expectation of 8797.0 events, divided as follows: 7030.1 (79.9\%)
from the \twonu\ decay of $^{76}$Ge; 1244.6 (14.1\%) from $^{42}$K; 335.5 (3.8\%) 
from $^{214}$Bi; and 186.8 (2.1\%) from $^{40}$K. The individual 
components derived from the fit are also shown in figure~\ref{fig:model}. 
The signal-to-background ratio in the region 600$-$1800~keV is on 
average $4:1$, which is much better than for any past experiment which 
observed the \twonu\ decay of $^{76}$Ge. The best ratio achieved so far 
for \gess\ was approximately $1:1$ by \hdm~\cite{klap01}. 
The model is able to reproduce well the experimental data, 
as shown in the lower panel of figure~\ref{fig:model}. The $p$-value of the fit  
derived from the procedure of \cite{allen2}, is $p=0.77$.
All nuisance parameters are eventually integrated over in order to derive the
posterior pdf $p(\thalf)$ for the \twonu\ half-life. The $p(\thalf)$ distribution is 
nearly Gaussian; the best estimate of the half-life is 
\begin{equation}
\thalf \ = \ (1.84 ^{+0.09}_{-0.08}) \cdot 10^{21} \ {\rm yr,} 
\end{equation}
(fit error only). This uncertainty is calculated as  
the smallest interval containing 68\% probability of $p(\thalf)$. It 
includes the uncertainty induced on $\thalf$ by the nuisance parameters 
of the fit and accounts for parameter correlations. 
Active masses and \gess\ isotopic abundances drive the fit uncertainty 
on $\thalf$: if these parameters were known without uncertainties, the $1 \sigma$ 
uncertainty from the fit would be about $0.03 \cdot 10^{21}$~yr.\\
\begin{table}[tbp]
\begin{center}
\begin{tabular}{l | c c c |c}
detector & total mass & active mass & \gess\ isotopic & $\thalf$ \\
 & (g) & (g) & abundance (\%) & ($10^{21}$~yr) \\
\hline
ANG2 & 2833 & 2468$\pm$121$\pm$89 & 86.6$\pm$2.5 & $1.99^{+0.14}_{-0.15}$ \\
ANG3 & 2391 & 2070$\pm$118$\pm$77 & 88.3$\pm$2.6 & $1.69^{+0.15}_{-0.14}$\\
ANG4 & 2372 & 2136$\pm$116$\pm$79 & 86.3$\pm$1.3 & $1.94^{+0.14}_{-0.15}$\\
ANG5 & 2746 & 2281$\pm$109$\pm$82 & 85.6$\pm$1.3 & $1.79^{+0.12}_{-0.14}$ \\
RG1 & 2110 & 1908$\pm$109$\pm$72 & 85.5$\pm$2.0 & $1.94^{+0.18}_{-0.14}$\\
RG2 & 2166 & 1800$\pm$99$\pm$65 & 85.5$\pm$2.0 & $1.93^{+0.16}_{-0.16}$\\
\end{tabular}
\end{center}
\caption{Summary table of the six enriched detectors used in this work. For each 
detector, the total mass, the active mass 
and the isotopic abundance of $^{76}$Ge are given. 
The uncertainties reported for the active masses are the uncorrelated (first) 
and correlated (second) errors. 
The last column reports the $\thalf$ which is obtained by performing the analysis 
on each detector individually. The uncertainty on $\thalf$ is 
the fit uncertainty only.
}\label{tab:detectors}
\end{table}

As a cross-check, the same procedure is run for each detector 
separately. The resulting $\thalf$ values are summarized 
in table~\ref{tab:detectors}; they are mutually consistent 
within their uncertainties ($\chi^2$/$\nu$ = 3.02/5).

\subsection{Systematic uncertainties} \label{sec:syst}
\begin{table}[tbp]
\begin{center}
\begin{tabular}{l | cc}
item & \multicolumn{2}{c}{uncertainty on $\thalf$} \\
 &  \multicolumn{2}{c}{(\%)} \\
\hline
non-identified background components & $+ 5.3$ &  \\
energy spectra from $^{42}$K, $^{40}$K and $^{214}$Bi & $\pm 2.1$ & \\
shape of the \twonu\ decay spectrum & $\pm 1$ & \\
\hline
subtotal fit model &  & $^{+5.8}_{-2.3} $ \\
\hline
precision of the Monte Carlo geometry model & $\pm 1$ & \\
accuracy of the Monte Carlo tracking & $\pm 2$ & \\
\hline
subtotal Monte Carlo & & $\pm 2.2$ \\
\hline
data acquisition and selection & & $\pm 0.5$ \\
\hline
Grand Total & & $^{+6.2}_{-3.3} $ \\
\end{tabular}
\end{center}
\caption{Summary table of the systematic uncertainties on $\thalf$ which are taken into 
account for this work and which are not included in the fitting procedure.}\label{tab:syst}
\end{table}

The items which are taken into account as possible systematic 
uncertainties of $\thalf$ and which are not included in the fitting procedure 
are summarized in table~\ref{tab:syst}. They can be 
divided into three main categories: (1) uncertainties related 
to the fit  model (choice of the components, shape of input spectra); (2) uncertainties 
due to the Monte Carlo simulation regarding the precision of the 
geometry model and the accuracy of the tracking of particles; 
(3) uncertainties due to data acquisition and handling. The latter term 
turns out to be negligible with respect to the others. The most relevant 
items from table~\ref{tab:syst} are briefly discussed in the following. \\

Additional background components that are not accounted for in the fit model 
might be present in the \gerda\ spectrum (see ref.~\cite{NIM_paper} for a list 
of the $\gamma$-ray lines detected in the \gerda\ spectrum and of the corresponding 
intensities). 
Due to the large signal-to-background ratio and the limited exposure
these background components cannot be identified unambiguously.
The uncertainty arising from such possible contributions is estimated to 
be $+5.3\%$. Since any further background component would lead to 
a longer $\thalf$, this uncertainty is asymmetric. 
It is estimated by performing a fit with the contributions from 
$^{60}$Co, $^{228}$Ac, and a flat background added to the model. 
These additional components are treated in the same way as the ``standard" 
background components ($^{42}$K, $^{40}$K, and $^{214}$Bi). 
The spectra from $^{60}$Co and $^{228}$Ac are simulated by Monte Carlo 
assuming close sources and one additional parameter for each detector 
and each additional background contribution is included in the fit. 
Also for the flat background an individual contribution is considered for 
each detector. The flat component describes the contribution coming 
from $^{208}$Tl decays from the $^{232}$Th chain: given the small number 
of events expected in the analysis energy window, this contribution can 
be roughly approximated to be constant. 
To a first approximation, also other possible non-identified 
background components can be accounted by the constant contribution to the 
model. \\

The systematic uncertainty on $\thalf$ due to the uncertainties in the 
spectra of the standard background components ($^{42}$K, $^{40}$K, and $^{214}$Bi) 
is estimated to be 2.1\%. It is evaluated by repeating the analysis with 
different assumptions on the position and distribution of the sources and 
with artificial variations (e.g. via a scaling factor) of the ratio between the 
full-energy peaks and the Compton continua.\\

The primary spectrum of the \twonu\ decay which is fed 
into the Monte Carlo simulation is generated by the code \textsc{decay0}. 
\textsc{Decay0} implements the algorithm described in \cite{tre95}, which 
is based on \cite{papers2nu,blatt63}.
The \twonu\ decay distributions of \cite{tre95} are in 
principle more precise than those based on the Primakoff-Rosen 
approximation~\cite{pr}. They 
have been cross-checked against the high-statistics data of the \textsc{Nemo} 
experiment for several nuclei: $^{82}$Se, $^{96}$Zn and 
$^{150}$Nd~\cite{nemo}. The \twonu\ spectrum 
derived by the Primakoff-Rosen 
approximation was used in earlier works 
with $^{76}$Ge, like \cite{klap03}. 
When the present analysis is re-run by 
using the formula of \cite{klap03}, the $\thalf$ result is 
stable within 1\%. \\

The uncertainty related to the \textsc{MaGe} Monte Carlo simulation 
arises from two sources: (1) the implementation of the experimental geometry into 
the code (dimensions, displacements, materials); and (2) the interaction of 
radiation with matter (cross sections, final state distributions) which is 
performed by \textsc{Geant4}. These items are evaluated to be 1\% and 2\%, 
respectively. The estimated contribution due to the particle tracking is 
based on the fact that electromagnetic physics processes provided by \textsc{Geant4} 
for $\gamma$-rays and e$^\pm$ have been systematically validated 
at the few-percent level in the energy range which is relevant for $\gamma$-ray 
spectroscopy~\cite{nist}. In this particular application 
the Monte Carlo uncertainty is mainly due to the propagation 
of the external $\gamma$-rays: the \twonu-decay electrons generated in the 
germanium detectors have a sub-cm range and they usually deposit their entire 
kinetic energy, apart from small losses due to the escape of Bremsstrahlung or 
fluorescence photons. \\

The combination in quadrature of the contributions reported in table~\ref{tab:syst} 
sums up to $^{+6.2}_{-3.3} \%$, corresponding to $^{+0.11}_{-0.06} \cdot 10^{21}$~yr.

\section{Results and conclusions}
The half-life of the \twonu\ decay of $^{76}$Ge was derived from 
the first data from the \gerda\ experiment at the INFN Gran Sasso Laboratory. 
\textsc{Gerda} operates HPGe detectors enriched in \gess\ directly immersed in 
liquid argon. The analysis has been carried out on the data collected with six 
\geenr\ detectors (14.6~kg total weight) during 125.9~live days 
by fitting the energy spectra with a comprehensive model. The best estimate of 
the half-life of the \twonu\ decay is
\begin{equation}
\thalf \ = \ (1.84 ^{+0.09}_{-0.08 \ {\rm fit}}\, 
^{+0.11}_{-0.06 \ {\rm syst}}) \cdot 10^{21} 
\ {\rm yr} \ = \ (1.84 ^{+0.14}_{-0.10}) \cdot 10^{21} \ {\rm yr,}
\end{equation}
with the fit and systematic uncertainties combined in quadrature.\\
\begin{figure}[tpb]
\begin{center}
\includegraphics[width=0.85\columnwidth]{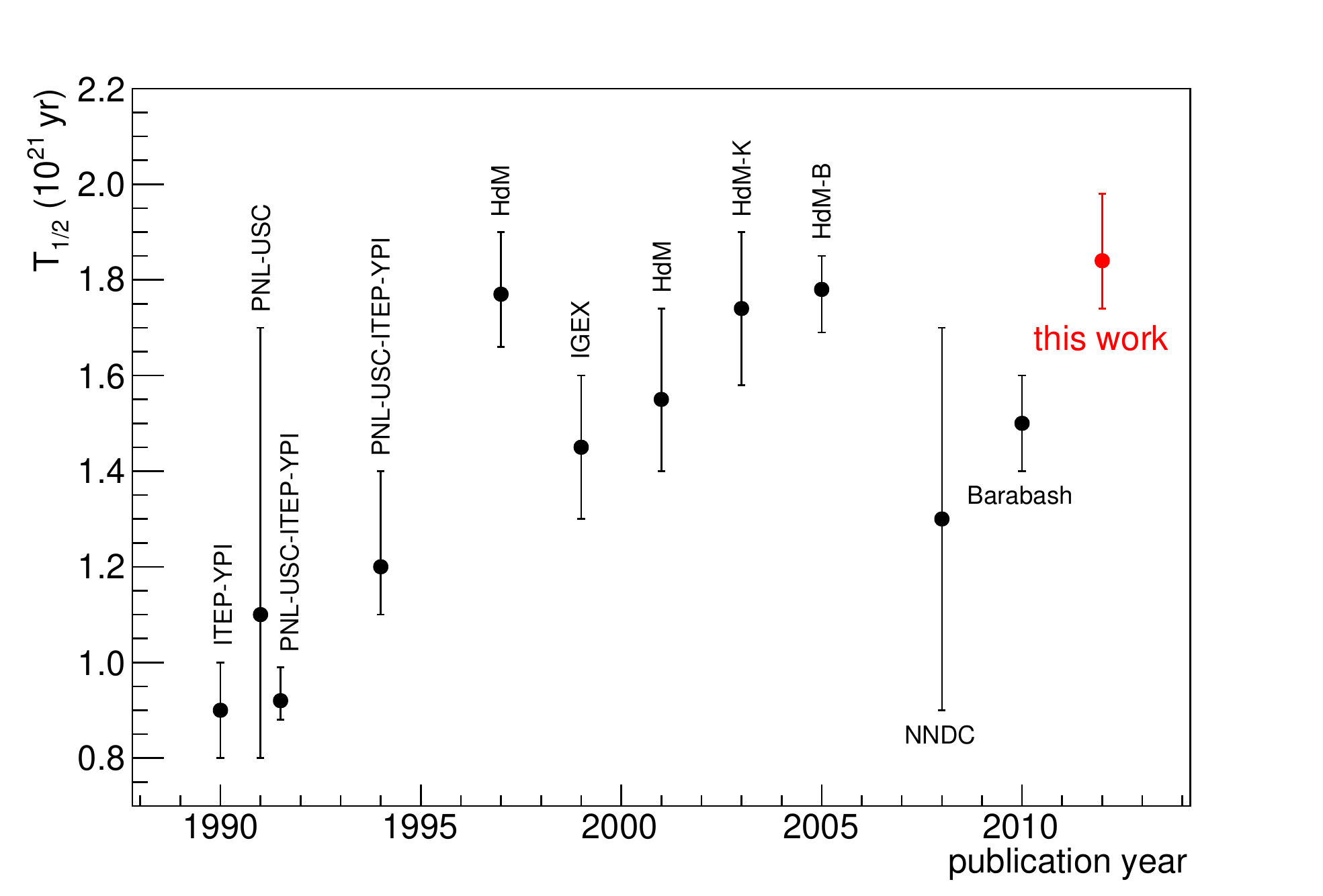}
\caption{Experimental results for $\thalf$ of $^{76}$Ge vs. publication 
year. The plot includes results from the experiments 
ITEP-YPI~\cite{vas90}, PNL-USC (\genat)~\cite{mil90} 
PNL-USC-ITEP-YPI~\cite{avi91,avi94}, 
Heidelberg-Moscow (\hdm)~\cite{gun97,klap01} and \igex~\cite{igex99,igex03}, as well as 
the re-analysis of the \hdm\ data by Klapdor-Kleingrothaus \etal\ \cite{klap03} (HdM-K) 
and by Bakalyarov \etal\ \cite{bak03} (HdM-B). The NNDC-recommended value~\cite{prit08} 
and the global weighted average evaluated by Barabash~\cite{bar10} are 
also shown.}\label{fig:pubhistory}
\end{center}
\end{figure}

The half-life is longer than all previous measurements reported in the literature. A summary 
of the $\thalf$ results of $^{76}$Ge from earlier experiments vs. 
publication year is displayed in figure~\ref{fig:pubhistory}. The figure includes 
nine measurements published between 1990 and 
2005 and two weighted averages. A trend towards longer $\thalf$ values 
reported by the most 
recent (and lowest background) experiments is clearly seen. The \gerda\ result 
is in better agreement with the two most recent results of \cite{klap03,bak03} 
that are based on the re-analysis of the \hdm\ data\footnote{If
$\thalf$ were as short as $1.5 \cdot 10^{21}$~yr reported in \cite{bar10}, 
the \twonu\ decay of $^{76}$Ge would account for nearly all counts detected in the range
600--1800 keV (expected: 8782.7, detected: 8976), thus leaving almost no
space for any other background source in \gerda.}. The fact that the half-lives  
derived in the more recent works $-$ and particularly in this one $-$ 
are systematically longer is probably related to the superior signal-to-background 
ratio, which lessens the relevance of the background modelling and subtraction.  
Thus the total \gerda\ uncertainty is comparable to what was achieved by 
\hdm, in spite of the much smaller exposure. \\
The uncertainty of the \gerda\ result can be further reduced in the future by
accumulating more exposure 
and by performing a new and more precise measurement of the active mass of 
the detectors. Large data sets will reduce also the 
uncertainty due to the fit model, as background components can be 
better characterized and constrained by the (non-)observation of $\gamma$ lines. \\ 

Using phase space factors from the improved electron wave functions reported in 
\cite{phasespace}, the experimental matrix element for the \twonu\ decay of 
\gess\ calculated with the half-life of this work is 
NME$^{2\nu} = 0.133^{+0.004}_{-0.005}$~MeV$^{-1}$. \\
The present value for NME$^{2\nu}$ is 11~\% smaller than that used in
\cite{rod06}, which compares the matrix elements
for \twonu\ and $0\nu\beta\beta$ decays. 
Ref.~\cite{rod06} also shows a relation between the 
two matrix elements for QRPA calculations of \gess. According to this, 
the new value for NME$^{2\nu}$  results in an increase of the 
predicted half-life for the $0\nu\beta\beta$ by about 15\%, which is well within 
the present uncertainty of the model calculation.\\
The nuclear matrix elements of the \twonu\ decay of \gess\ estimated 
from the charge exchange reactions (d,$^2$He) and 
($^3$He,t) are 
$(0.159 \pm 0.023)$ MeV$^{-1}$ \cite{kvi} and 
$(0.23 \pm 0.07)$ MeV$^{-1}$ \cite{rcnp}, respectively. They both seem to be higher 
than the value reported in this work, but consistent within the uncertainties.

\ack{The \gerda\ experiment is supported financially by 
the German Federal Ministry for Education and Research (BMBF),
the German Research Foundation (DFG) via the Excellence Cluster Universe,
the Italian Istituto Nazionale di Fisica Nucleare (INFN),
the Max Planck Society (MPG), the Polish National Science Centre (NCN),
the Russian Foundation for Basic Research (RFBR), and
the Swiss National Science Foundation (SNF).
The institutions acknowledge also internal financial support.\\
The \gerda\ collaboration thanks the Directors and the staff of the 
Laboratori Nazionali del Gran Sasso for their continuous strong support of the 
experiment.\\
We would like to thank Prof.~V.I.~Tretyak for help and suggestions 
about the modelling of the energy spectrum of the \twonu\ decay.}

\section*{References}





\end{document}